\documentclass[12pt]{article}
\usepackage{indentfirst,psfig}
\usepackage{cite}
\usepackage{doublespace}
\usepackage{citesupernumber}
\begin{document}

\title{Disorder-Induced Time-Dependent Diffusion in Zeolites}

\author{
Ligang Chen, Marco Falcioni, and Michael W. Deem\\
Department of Chemical Engineering\\
University of California\\
Los Angeles, CA  90095-1592}

\maketitle

\begin{abstract}
We suggest that disordered framework aluminums and non-framework
cations can create a disordered electrostatic potential in zeolites
that can lead to a discrepancy between diffusivities measured by
microscopic and macroscopic experimental techniques.  We calculate the
value of the discrepancy and the characteristic time scale at which it
occurs for several ionic and polarizable species diffusing in
zeolites.  For ionic species, a discrepancy is almost inevitable.  For
polarizable species, a significant discrepancy may occur
in some zeolites only for long alkanes or large species such as
benzene.
\end{abstract}

%\pacs{66.30.Jt, 05.60.Cd, 47.55.Mh}

\section{Introduction}
Disordered microporous materials, such as zeolites, microporous
carbons, pillared clays, and superionic conductors, have a wide range
of applications in fields such as catalysis, separation, remediation,
and sensing. In all of these uses, the dynamical behavior of the
sorbed molecules or ions can be a constraining factor. Due to the
complicated guest-host interactions in these systems, our ability to
predict the dynamical behavior on all time scales is still
limited. Accurate prediction of diffusivities is, however, an
important component in the design of practical processes using
microporous materials.

Among these materials, zeolites have been the focus of particularly
intense research efforts due to their wide industrial use and unique
crystal structures. Significant effort has been devoted to
understanding diffusion in particular.\cite{Karger} Despite this
study, a fundamental question persists for some classes of zeolites:
What causes the well-known discrepancy between the diffusivities
measured with macroscopic and microscopic techniques?  Macroscopic
methods, such as the zero-length column (ZLC) technique\cite{Eic} and the
Wicke-Kallenbach (WK) membrane method,\cite{Sun} typically measure
transport diffusivities under a concentration gradient. The time
scales involved in the macroscopic experiments are long ($\simeq
1\,$s).  Microscopic techniques, such as NMR, measure
self-diffusivities by following the motion of tagged molecules. In
this case the system is kept under equilibrium conditions. The NMR
technique samples motion on the \AA{ngstrom} length scale and probes
relatively short time scales ($\simeq 10^{-12}$--$10^{-8}\,$s) for the
diffusivities typical in zeolites. Mesoscopic methods, such as NMR
pulsed field gradient (PFG)\cite{Karger2} and quasi-elastic neutron
scattering (QENS),\cite{Jobic} also follow the microscopic motion of
diffusing species. These methods are time-scale limited, able to
access the $10^{-11}-10^{-3}\,$s range. The diffusivities measured
using macroscopic techniques can disagree from those measured by
microscopic techniques by one or more orders of magnitude.  For
example, the diffusivity of benzene in NaX crystals at $403$K is found
to be $1\times 10^{-10}\,$m$^2$s$^{-1}$ by NMR, but $4.5\times
10^{-12}\, $m$^2$s$^{-1}$ by ZLC.\cite{Karger:p515}

Several explanations for the discrepancy between macroscopic and
microscopic diffusivities have been offered, including an
extracrystalline resistance, surface barriers to mass transfer, and
internal heat conduction effects.\cite{Karger} To date, however, there
is no general theory that can satisfactorily explain the
discrepancy. Recent computer simulations have begun to shed light on
the understanding of this long-standing puzzle. It appears that
different types of molecular motions happen at different time
scales. Molecular dynamics simulations by Maginn {\it et al.}, for
example, reveal that the time constant associated with molecular
rearrangements of alkanes may be greater than the time constant for
motion along a given channel.\cite{Runnebaum} Similarly, kinetic Monte
Carlo simulation studies of benzene in zeolites by Auerbach {\it et
al.} suggest that, due to the different experimental time scales, PFG
NMR measures the intercage diffusivity, while the NMR relaxation
method measures the intracage diffusivity.\cite{Auerbach}

What these simulation studies suggest, and indeed what the discrepancy
between microscopic and macroscopic diffusivities suggests, is that
the diffusivity is actually time dependent.  That is, for some
sorbates in some zeolites, there seems to be one value of the
diffusivity, $D_0$, associated with motion on the \AA{ngstrom} scale
and another, smaller, value, $D_{\infty}$, associated with motions on
longer spatial scales.  Such a time-dependent diffusivity can arise
from an unusual time dependence of the mean-square displacement of the
diffusing species.  In this paper, we provide a theory to calculate
the diffusivity on different time and length scales.  This theory
gives not only the ratio $D_{\infty}/D_0$, but also the characteristic
time and length scale on which the measured diffusivity changes most
dramatically.

Although not generally appreciated, disorder within the zeolite
framework can have a dramatic effect on the measured diffusivity.  The
aluminum framework species as well as the associated, non-framework
cations create a disordered electrostatic environment for diffusing
species if the aluminums or cations are randomly sited.  The strength
of the randomness is determined by the aluminum loading, which can
vary from a 1:1 Si:Al ratio to a 3000:1 or higher Si:Al ratio. Higher
levels of aluminum loading require higher densities of exchangeable
cations.\cite{Newsam} Together, the framework aluminum and
non-framework cations create a random environment and lead to a lower
ratio of $D_{\infty}/D_0$.\cite{Deem} On the other hand, the disorder
can either increase or decrease the microscopic diffusivity $D_0$.  In
the absence of any specific interactions between the diffusing species
and the disordered ions, one would expect $D_0$ to decrease with
increasing disorder. For chemisorption-type interactions typical of
nucleophilic species, $D_0$ may increase with an increasing density of
cationic sites since the hopping distance between the sites decreases,
yet the binding energy per site may remain constant.\cite{Auerbach2}

In this paper, we model the long-distance features of the random
electrostatic potential in the zeolite with a particularly simple,
Gaussian form.  We analyze this model to give an explicit form of the
time-dependent, experimental diffusivity $D(t) / D_0$.  With
this result we can predict both the ratio of the macroscopic to the
microscopic diffusivity and the time or length scale at which
the difference between the two values becomes apparent.  If this time
scale is greater than that accessed by, say, the NMR technique, then
disorder provides one possible mechanism for a discrepancy between
microscopic and macroscopic diffusion measurements.

To perform this analysis, we use a field theoretic formulation of the
diffusion process. While the theoretical treatment is involved, the
physical mechanism considered is simply the slowing down of a
diffusing species by the random electric fields within the zeolite. Our
paper is organized as follows. In Section~\ref{sec:rg} we use
renormalization group theory to derive the time-dependent diffusivity
in terms of the disordered environment within the zeolite. The long-time
renormalization group result is matched to a short-time perturbation
theory to give an explicit prediction for the diffusivity. In
Section~\ref{sec:num} we describe Monte Carlo simulations that
validate the time-dependent theory.  Several forms of random disorder
are studied, and some numerical issues are discussed. In
Section~\ref{sec:zeo} we apply the theory to zeolites.  The reader
interested primarily in how disorder may explain the discrepancy
between microscopic and macroscopic measurements of the diffusivity
may wish to skip directly to this section.  Diffusion of both ionic
and polarizable species is considered in this section.  Values of the
discrepancy between microscopic and macroscopic measurements are
predicted in several cases, and the characteristic time and length
scales when the discrepancy becomes apparent are also given.  For
measurements of ionic diffusivities, discrepancies are almost
inevitable.  For polarizable species, discrepancies can be one order of
magnitude or more for long alkanes or large species such as benzene.
In Section~\ref{sec:last}, we summarize our findings and make some
final remarks.

\section{Time-Dependent Renormalization Group Theory}
\label{sec:rg}

Before specializing to the case of diffusion of ions or polarizable
species at low densities in disordered zeolites, we first consider the
more general problem of diffusion of any type of species in a
disordered environment.  In general, the disorder will slow down the
diffusing species, and we would like to predict the magnitude of this
effect and the characteristic time scale at which it occurs.  Since we
will be interested only in the ratio of the macroscopic diffusivity to
the microscopic diffusivity, we can use a theory valid for long times
and large distances.  Field theory is just such a tool.  Field theory
also naturally allows us to consider the effect of the disordered
environment. Alternatively, field theory can be viewed as one
particularly effective way to solve for the Green function of this
diffusion problem.  Interactions between the diffusing species will be
negligible at low densities, even for ionic species.\cite{Deem3,Deem4} We,
therefore, consider the single-species diffusion equation
\begin{equation}
\frac{\partial G_v({\bf x},t)}{\partial t} =
D_0 \nabla^2 G_v({\bf x}, t) + \beta D_0 \nabla \cdot
[G_v({\bf x}, t) \nabla v({\bf x})] 
\label{1}
\end{equation}
Here $v$ is the quenched random energy field caused by the disorder,
$G_v$ is the Green function of a single diffusing species in a given
realization of the disorder, $D_0$ is the ``bare,'' short-time
diffusivity, and $\beta=1/(k_{\rm B}T)$ is the inverse temperature.
The mean-square displacement is given by $R_v^2(t) = \int d^d {\bf
x}~\vert {\bf x} \vert^2 G_v({\bf x},t)$.  The mean-square
displacement measured in an experiment is averaged over the statistics
of the disorder: $R^2(t) = \langle R_v^2(t) \rangle =\int d^d {\bf
x}~\vert {\bf x} \vert^2 G({\bf x},t)$, where the Green function averaged
over disorder is $G({\bf x}, t)=\langle G_v({\bf x}, t) \rangle$.

Because we are interested only in how the large-distance features of
the random environment affect the diffusive motion, we can use a
particularly simple, Gaussian form for the disordered energy field.
Other forms turn out to be equivalent to this choice in a certain,
technical sense.  Since the force on a diffusing species is
proportional to the gradient of the energy field, we can set the average
value of the field to zero without loss of generality.  And since
we assume it is Gaussian, the only other quantity to
specify is the energy-energy correlation function $\chi_{v
v}({\bf r}) = \langle v({\bf 0}) v({\bf r}) \rangle$.

The time-dependent field theory and model we use has been analyzed
extensively in the context of reaction and diffusion processes in
random media.\cite{Deem2, Park} Here we consider only the diffusion
process.  In this section, we derive a formula for the time-dependent
self-diffusivity, which is defined as
\begin{equation}
D(t) = \frac{R^2(t)}{2 d t}
\label{eq2}
\end{equation}
For the application to zeolites we will consider $d=3$ dimensions, but
here we develop the theory for a general spatial dimension.

The derivation of the field theory is somewhat technical,\cite{Deem2, Park}
and so we provide a summary of the procedure here.  A 
master equation that describes the motion of the diffusing species
is first written.  This master equation is equivalent to eq \ref{1}.
The master equation is then rewritten in terms of creation and
annihilation operators.  The coherent state representation is then
used to rewrite the master equation in terms of a field theory.
Finally, the replica trick is used to integrate out the effects of
the disorder.  After these exact transformations, one arrives at the
action $S=S_0+S_I$, with
\begin{eqnarray}
S_0 &=& \sum_{\alpha = 1}^{N} \int d^d {\bf x} \int_0^{\infty}
        dt~{\bar a}_{\alpha}({\bf x}, t) [\partial_t - D_0\nabla^2 +
        \delta(t)]a_{\alpha}({\bf x}, t) \\ \nonumber 
S_I &=& \frac{\beta^2 D^2_0}{2}\sum_{\alpha, \gamma =1 }^{N}
        \int_0^{\infty} dt_1 dt_2 \int_{{\bf k}_1 {\bf k}_2 {\bf k}_3
        {\bf k}_4} (2\pi)^d \delta({{\bf k}_1+{\bf k}_2+{\bf k}_3+{\bf
        k}_4}) \nonumber \\
    & & \hat{\bar{a}}_{\alpha}({\bf k}_1, t_1) \hat{a}_{\alpha} ({\bf
        k}_2, t_1) \hat{\bar{a}}_{\gamma}({\bf k}_3, t_2)
        \hat{a}_{\gamma}({\bf k}_4, t_2) \nonumber \\ & & {\bf k}_1
        \cdot ({\bf k}_1+{\bf k}_2) {\bf k}_3 \cdot ({\bf k}_1+{\bf
        k}_2) \hat{\chi}_{vv}(|{\bf k}_1+{\bf k}_2|)
\end{eqnarray}
The Fourier transform of a function is defined by $\hat{f}({\bf
k})=\int d^d {\bf x}~f({\bf x}) \exp(i{\bf k} \cdot {\bf x})$, and the
notation $\int_{\bf k}$ is a short hand for $\int d^d{\bf
k}/(2\pi)^d$.   The number of replicas is $N$, and the $N \to 0$ limit
is implied.
The action is used to calculate averages.
 The Green function, for example, is given by
\begin{equation}
G({\bf x}, t) = \lim_{N \to 0}
\frac{1}{N} \sum_{i=1}^N \langle a_i({\bf x}, t) \bar a_i({\bf 0}, 0) \rangle
\label{2b}
\end{equation}
where the average is with respect to the weight $\exp(-S)$. 
The free theory term $S_0$ represents simple
diffusion. The interaction term $S_I$ comes from averaging the
diffusion process over the random energy field, which is again assumed to be
Gaussian, with zero mean and correlation function
$\hat{\chi}_{vv}$. The form of $\hat{\chi}_{vv}$ is left general
throughout the calculation.

Using field theoretic renormalization group theory\cite{Justin} we
let $ \bar{a} \rightarrow Z_a^{\frac{1}{2}} \bar{a},\ a \rightarrow
Z_a^{\frac{1}{2}} a,\ D \rightarrow Z_a^{-1} Z_D D,$ and $\beta^2
\rightarrow \beta^2 Z_D^{-2} Z_I.$ Note that the dimensions of
$\beta^2 \hat{\chi}_{vv}$ are the same as that of $k^{-d}$, where $k$ has
the dimensions of wave number. We have chosen $\beta$ as dimensionless
in this renormalization scheme. The $Z$ factors will be determined by
a subtraction scheme. Following a one-loop calculation of the
renormalized vertex functions in the long-time limit, we find 
\begin{eqnarray}
Z_a &=& 1 \nonumber \\
Z_D &=& 1 - \frac{\beta^2 K_d}{d} \int_{\mu}^{\Lambda} dk\ k^{d-1}
	\hat{\chi}_{vv}(k) \nonumber \\ 
Z_I &=&  1 - \frac{2\beta^2 K_d}{d} \int_{\mu}^{\Lambda} dk\ k^{d-1}
	\hat{\chi}_{vv}(k)
\end{eqnarray} 
where $K_d=S_d/(2\pi)^d$, $S_d=2\pi^{d/2}/[\Gamma(d/2)]$ is the
surface area in $d$ dimensions, $\Lambda$ is a cut-off, and
$\mu=\Lambda/e^l$ with $l$ the flow parameter. No contribution
renormalizes the time derivative or delta function term in $S_0$. The
associated renormalization-group functions are defined by
\begin{equation}
\gamma_i=\mu \left. \frac{\partial \ln Z_i}{\partial
\mu}\right|_{0},\ i=a, D, I
\end{equation}
where the subscript 0 on the derivative indicates fixed bare
variables. We find
\begin{eqnarray}
\gamma_a &=&0 \nonumber \\
\gamma_D &=&\beta^2 K_d \mu^d \hat{\chi}_{vv}(\mu)/d \nonumber \\ 
\gamma_I &=&2\beta^2 K_d \mu^d \hat{\chi}_{vv}(\mu)/d
\end{eqnarray}
and
\begin{equation}
\beta_I=\mu \left. \frac{\partial \beta^2}{\partial
\mu}\right|_{0} =  (\gamma_I - 2 \gamma_D ) \beta^2 = 0
\end{equation} 
The dynamical exponent is given by
\begin{equation}
z = 2-\gamma_a+\gamma_D = 2+\beta^2 K_d \mu^d \hat{\chi}_{vv}
(\mu)/d\label{z1}
\end{equation}
The renormalized time flows according to
\begin{equation}
t(l^{*})=t\exp \left[-\int_0^{l^*} z(l)dl \right]=t_0 \label{z2}
\end{equation}
The matching time, $t_0$, is usually on the order of $1/(\Lambda^2
D_0)$, which is within the range of validity of both renormalization
group scaling and mean field theory. The Callan-Symanzik equation then
reads
\begin{equation}
\left(\mu \frac{\partial}{\partial \mu} - \gamma_D \right)D=0
\end{equation}
Solving this differential equation, we find
\begin{equation}
D(l^*) = D(0)\exp\left[-\frac{\beta^2 K_d}{d} \int_0^{l^*} dl
\left(\frac{\Lambda}{e^l} \right)^d \hat{\chi}_{vv} \left(\frac{\Lambda}
{e^l} \right) \right] \label{Dl}
\end{equation}
An explicit form for $D(t)$ follows from eq~\ref{Dl} after relating the
flow parameter $l^*$ to time by eqs~\ref{z1} and \ref{z2}
\begin{eqnarray}
D(t) &=& D_0\exp\left[-\frac{\beta^2 K_d}{d} \int_0^{l^*(t)} dl
         \left(\frac{\Lambda}{e^l} \right)^d \hat{\chi}_{vv}
         \left(\frac{\Lambda} {e^l} \right) \right] \nonumber \\ 
     &=& D_0\exp \left[-\frac{\beta^2 K_d}{d} \int_{\Lambda/e^{l^*}}
         ^{\Lambda} dk\, k^{d-1} \hat{\chi}_{vv}(k) \right]
\label{DRG}
\end{eqnarray}
In the limit of $t \rightarrow \infty$, which implies $l^* \rightarrow
\infty$, the above equation yields
\begin{equation}
D_{\infty}=D_0\exp[-\beta^2 \chi_{vv}(0)/d] \label{D_infty}
\end{equation} 
which is in agreement with the long-time renormalization group result
found in Refs \citen{Deem} and \citen{Dean}.
Finally, eq \ref{z2} can be rewritten as
\begin{equation}
t = t_0 e^{2l^{*}} D_0 / D(t)
 \label{tm}
\end{equation}

The renormalization result depends on the matching time $t_0$
appearing in eq \ref{z2}. We determine $t_0$ by equating perturbation
theory to the renormalization group result for weak disorder, or
equivalently, $\beta \rightarrow 0$. The mean square displacement is
given by $R^2(t) = 2dDt$, so the diffusivity is related to the Green
function by
\begin{equation}
D(t) = \frac{\int d^d{\bf x}~x^2 G({\bf x}, t)}{2dt} = -\frac
{\nabla^2_{\bf k} \hat{G}({\bf k}, t)|_{k=0}}{2dt}
\end{equation}
For weak disorder, perturbation theory gives
\begin{equation}
D(t)=D_0-\frac{\beta^2 K_d}{dt} \int_0^{\Lambda} dk\ k^{d-3} \hat{\chi}_{vv}(k)
(-3+x+3e^{-x}+2xe^{-x}) + O(\beta^4, t^3) \label{DPT}
\end{equation}
where $x \equiv D_0 k^2 t$. In the limit $\beta \rightarrow 0$ we
match perturbation theory and the renormalization group theory at a
characteristic time, $t_{\rm c}$:
\begin{equation}
D(t_{\rm c}) = (D_0 + D_{\infty})/2 = D_0-\frac{D_0 \beta^2 \chi_{vv}(0)}{2d}
+O(\beta^4) \label{Dtc}
\end{equation}
We find $t_{\rm c}$ by equating eqs \ref{DPT} and \ref{Dtc} to $O(\beta^2)$:
\begin{equation}
\chi_{vv}(0) = \frac{2 K_d}{D_0 t_{\rm c}} \int_0^{\Lambda} dk\ k^{d-3}
\hat{\chi}_{vv}(k) (-3+x_{\rm c}+3e^{-x_{\rm c}}+2x_{\rm c}e^{-x_{\rm
c}}) \label{matching}
\end{equation}
where $x_{\rm c}=D_0 k^2 t_{\rm c}$. Note that eq~\ref{matching} shows
that $t_{\rm c}$ is a function of $\Lambda$, $D_0$, and the form of
$\hat{\chi}_{vv}$, but not the strength of the disorder. We similarly
find $l_c$ by equating eqs \ref{DRG} and \ref{Dtc} to $O(\beta^2)$:
\begin{equation}
\chi_{vv}(0) = 2K_d \int_0^{l_c} dl \left(\frac{\Lambda}{e^l}\right)^d
\hat{\chi}_{vv} \left(\frac{\Lambda}{e^l} \right) \label{mrg}
\end{equation}
We finally obtain $t_0$ at small $\beta^2$ from the relation, $t_0 =
t_{\rm c}\ e^{-2 l_c}$. We assume that $t_0$ is only weakly dependent
on the strength of the disorder and use the same $t_0$ for all values
of $\beta^2$.

\section{Monte Carlo Simulation Results}
\label{sec:num}

Before we apply the time-dependent theory to diffusion in disordered
microporous materials, we first test it against computer simulations.
We do this by generating numerical solutions to eq \ref{1}, averaging
them over the statistics of the disorder, and then comparing the
results to eq \ref{DRG}.  The computer simulations are done on 
a lattice in two dimensions and follow closely the strategy of
Ref~\citen{Victor}. Using eqs (3-10) from Ref~\citen{Victor}, we
create a Gaussian random energy field, prescribed by $\hat{v}(k)$, on a
lattice.  We determine the energy field in real space by fast Fourier
transform. To obtain reasonable statistics for the diffusivity, we
follow the motion of 10000 independent particles on a lattice of size
$L$.  Periodic boundary conditions are enforced. We use lattices of
size $L=512$, 1024, and 2048 in order to identify and avoid finite
size effects. The initial particle positions are chosen at random.
Finally, $D(t)$ is measured by averaging $R^2(t)$ as a function of $t$ over a
moving window of 200 Monte Carlo steps and evaluating the slope. To
test the robustness of the theory, we examined three forms of
$\hat{\chi}_{vv}$: $\hat{\chi}_{vv}=\gamma \exp[-k^2/(2k_{\rm c}^2)]$,
$\hat{\chi}_{vv}=\gamma \exp(-\vert k\vert /k_{\rm c})$, and
$\hat{\chi}_{vv}=\gamma k_{\rm c}^4/(k^2 + k_{\rm c}^2)^2$.

\begin{figure}[htbp]
\centering
\leavevmode
\psfig{file=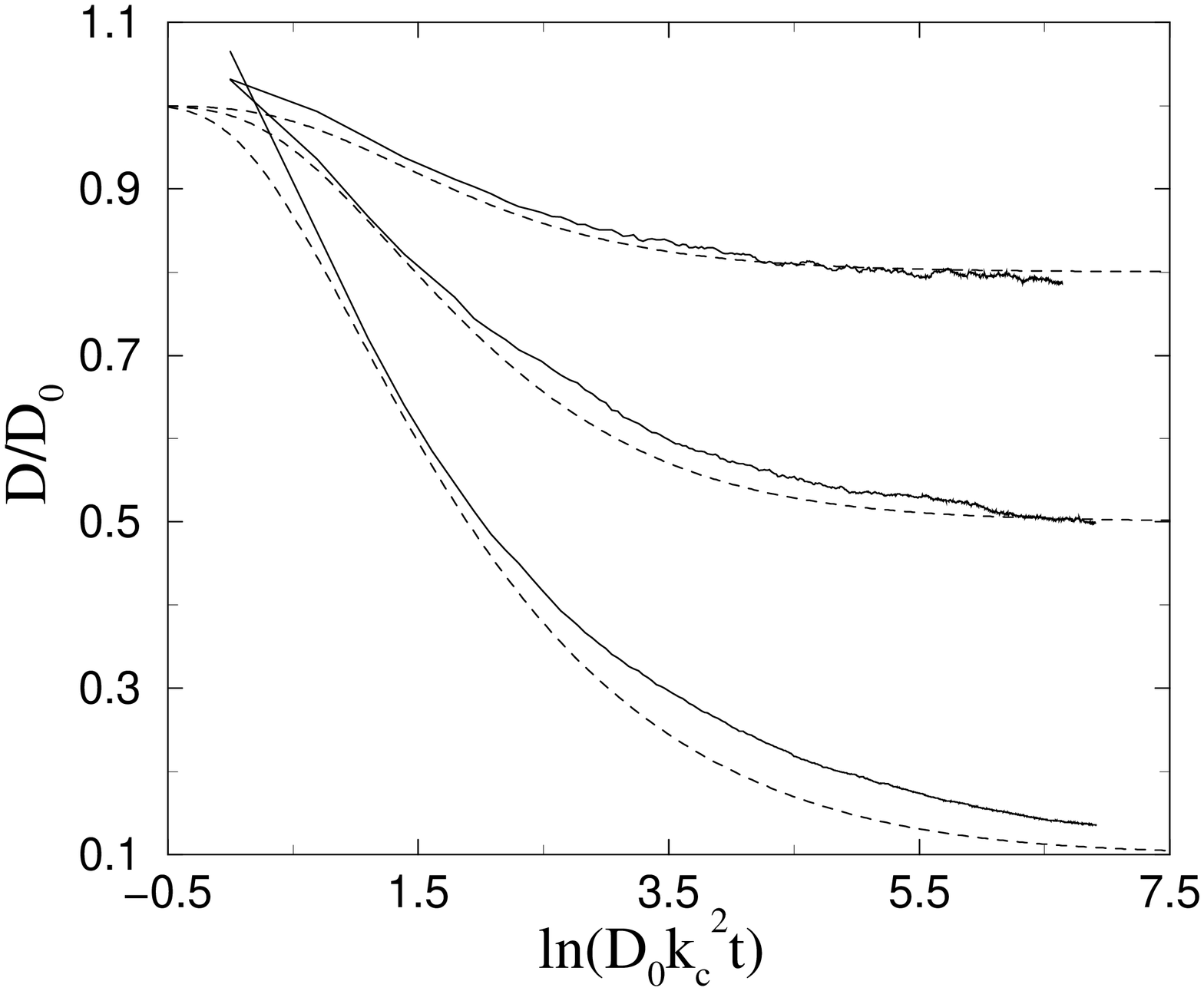,height=2in,angle=-90}
\caption{Comparison of the theoretical results (dashed) with
simulation data (solid) for $\hat{\chi}_{vv}(k)=\gamma
\exp[-k^2/(2k_{\rm c}^2)]$, with $k_{\rm c}=0.1$. Three different
values of $\gamma$ were used corresponding to $D_{\infty}/D_0=0.8,$
0.5, and 0.1, as predicted by eq \ref{D_infty}.}
\label{fig1}
\end{figure}

\begin{figure}[htbp]
\centering
\leavevmode
\psfig{file=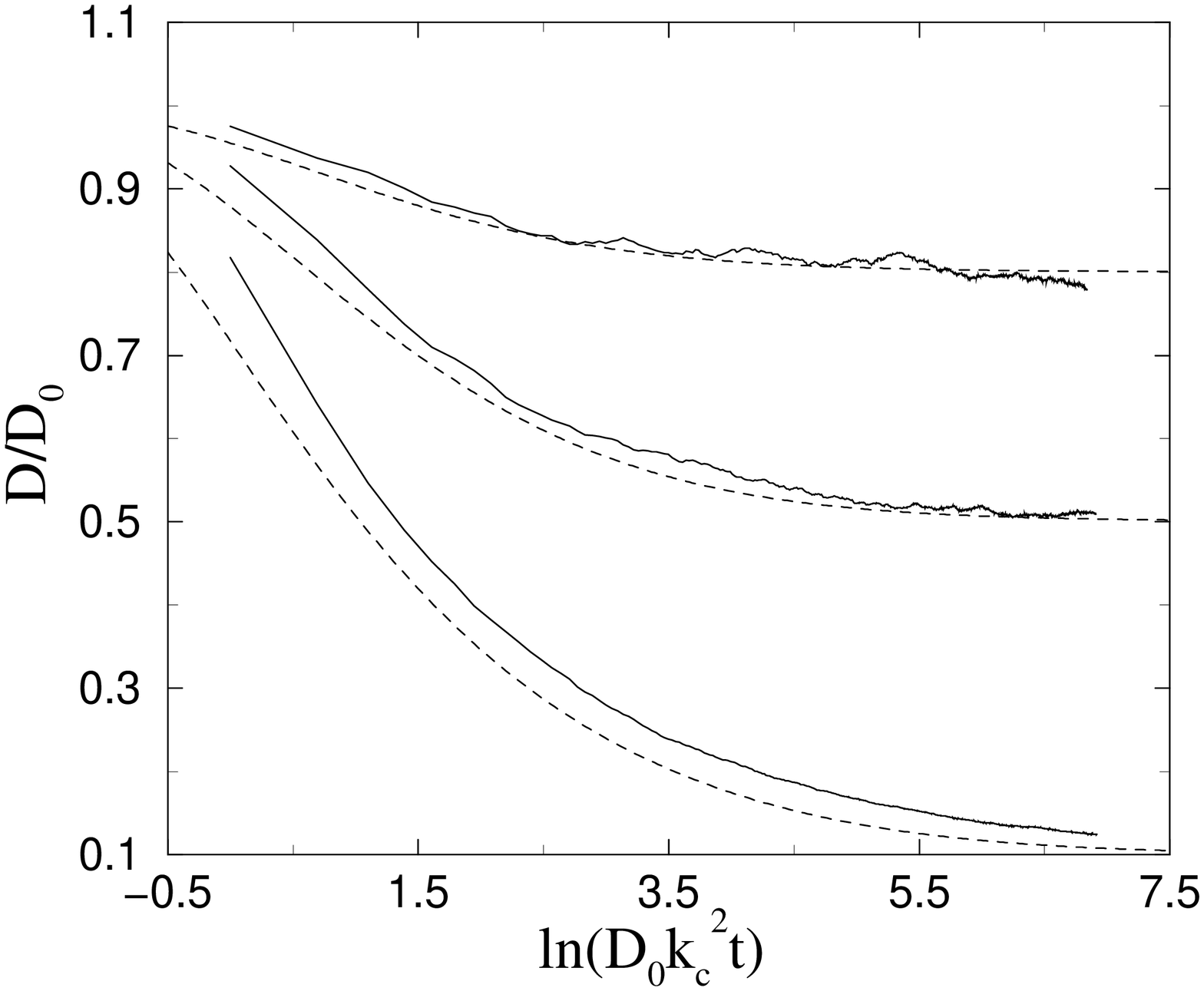,height=2in,angle=-90}
\caption{Comparison of the theoretical results (dashed) with
simulation data (solid) for $\hat{\chi}_{vv}(k)=\gamma \exp(-\vert
k\vert /k_{\rm c})$, with $k_{\rm c}=0.1$. Three different values of
$\gamma$ were used corresponding to $D_{\infty}/D_0=0.8,$ 0.5, and 0.1,
as predicted by eq \ref{D_infty}.}
\label{fig2}
\end{figure}

\begin{figure}[htbp]
\centering
\leavevmode
\psfig{file=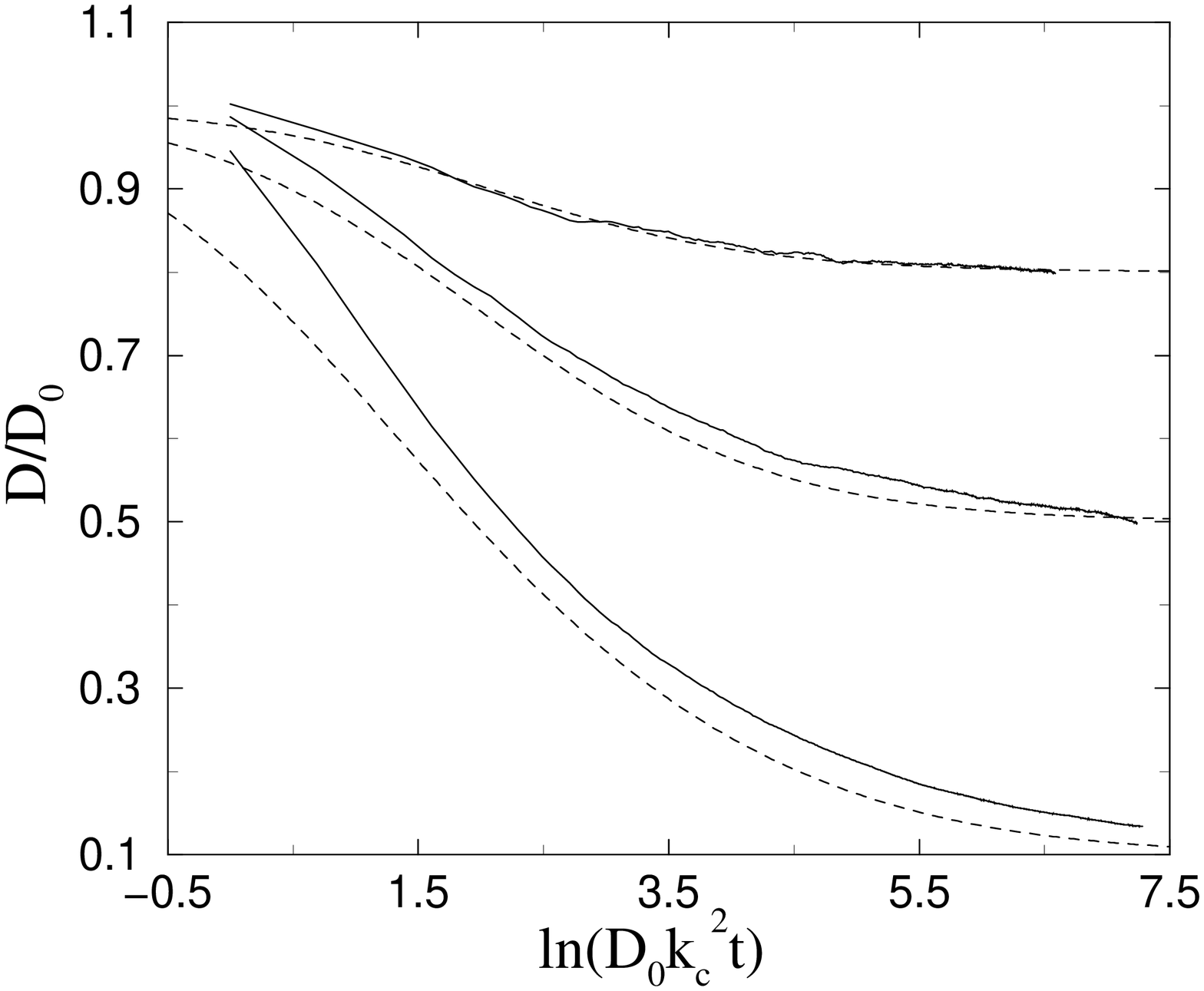,height=2in,angle=-90}
\caption{Comparison of the theoretical results (dashed) with
simulation data (solid) for $\hat{\chi}_{vv}(k)=\gamma k_{\rm
c}^4/(k^2 + k_{\rm c}^2)^2$, with $k_{\rm c}=0.1$. Three different
values of $\gamma$ were used corresponding to $D_{\infty}/D_0=0.8,$
0.5, and 0.1, as predicted by eq \ref{D_infty}.}
\label{fig3}
\end{figure}

The simulation results with $L=2048$ and $k_{\rm c}=0.1$ are shown in
Figure~\ref{fig1}-\ref{fig3} for each form of
$\hat{\chi}_{vv}$. Significantly, the infinite time result for the
diffusivity, eq~\ref{D_infty}, appears exact to within the statistics of
our simulation. More importantly for the present purpose, there is a
good agreement between the renormalization group result, eq~\ref{DRG},
and the simulation data over the entire time range. Indeed, $t_0$ appears
to be only weakly dependent on the strength of disorder. Recall that
$t_0$ was determined at weak disorder, and we assumed it to be
constant at strong disorder. The observed constancy of $t_0$ validates
our matching procedure.

Interestingly, two types of finite size effects can occur in these
simulations.  First, the lattice must be large enough to capture the
length scale at which the disorder begins to slow down the diffusing
species.  If the lattice is not large enough, the diffusivity will
reach the proper asymptotic, long-time value.
Second, the lattice must be fine enough so as to capture all of the
features inherent in the chosen $\chi_{vv}({\bf x})$.  If the
characteristic length of $\chi_{vv}({\bf x})$ is finer than the
lattice spacing, then some of the effects of the random energy field
will be lost.  Since the diffusion constant decreases with increasing
disorder, a lattice that is not fine enough and, therefore, neglects
some of the effects of the disorder, will predict a diffusivity that
is too large.  We, indeed, observe this effect for simulations with
$k_{\rm c} = 1.0$.  In this case, the characteristic size $r_{\rm c} =
2 \pi / k_{\rm c}$ is approaching the lattice spacing of unity that we
use.

\begin{figure}[htbp]
\centering
\leavevmode
\psfig{file=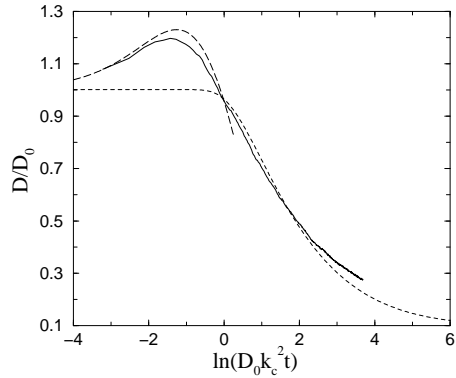,height=2in,angle=-90}
\caption{Comparison of the renormalization group (dashed) results with
simulation data (solid) and short-time perturbation theory
(long-dashed) for $\hat{\chi}_{vv}(k)=\gamma \exp[-k^2/(2k_{\rm
c}^2)]$, with $k_c=0.02$ and $D_{\infty}/D_0=0.1$.}
\label{fig4}
\end{figure}

Note that we observe a peak of $D(t)/D_0$ in the simulations for small
$D_0k_{\rm c}^2t$. This peak is more evident for small values of
$k_{\rm c}$. In Figure~\ref{fig4} we plot the simulation results
together with the renormalization group and short-time perturbation
theory results for $k_{\rm c}=0.02$.
 It is clear that perturbation theory describes the short-time
peak well. This is to be expected since the perturbation result is
accurate to $O(t^2)$ for all values of $\beta^2$.  In contrast, the
renormalization group approach fails to predict this interesting
short-time behavior, since the renormalization procedure is carried
out in the asymptotic, long-time limit. This peak in the diffusivity
is due to the rapid relaxation of the particles away from the initial,
non-equilibrium distribution. Consequently, this peak would not be
expected to be observed in equilibrium techniques such as NMR, PFG
NMR, or QENS.

\section{Time-Dependent Diffusion in Zeolites}
\label{sec:zeo}
Within a zeolite crystal there is a random potential field, mainly due
to the random electrostatic potential created by the charged framework
Al$^{-}$ and non-framework cation sites. The lattice defects, cations,
and associated aluminum atoms create strong disorder, increase the
hopping activation energy, inhibit the mobility of sorbates, and slow
down the diffusion process. We list the ionic self-diffusivities
measured in same-type of zeolites with different aluminum loadings in
Table \ref{ions}. To reduce the effects of ionic hydration, data taken
for methanol or ethanol solvents are exhibited. As can be seen, Table
\ref{ions} confirms a reduction diffusivity caused by
aluminum-induced disorder. That is, at higher aluminum loadings the
activation energy increases and the diffusivity decreases.

We assume that the spatial arrangement of the charges in zeolites is
fixed during formation conditions, i.e. that the disorder is
quenched.\cite{Ma} The Debye-H\"{u}ckel theory\cite{Netz} for the
density-density correlation function of the charged impurities then
gives
\begin{equation}
\hat{\chi}_{\rho \rho}(k) = \frac{\gamma k^2}{k^2+\kappa^2}
\end{equation}
with $ \gamma = |e^-|^2(\rho_{+} + \rho_{-})$, and $\kappa$ is an
inverse characteristic length of the charged disorder, defined by
$\kappa^2 = \beta_{\rm F} |e^-|^2(\rho_{+} + \rho_{-})/\epsilon_{\rm
F}$. Here $e^-$ is the electron charge, $\rho_{+}$ and $\rho_{-}$ are
the positive and negative charge number densities, respectively,
$\epsilon_{\rm F}$ is the dielectric constant under formation
conditions, and $\beta_{\rm F}=1/(k_BT_{\rm F})$ with $T_{\rm F}$ the
formation temperature. Under typical aqueous synthesis conditions of
${\rm pH}=12$ and $T_{\rm F}=398\,$K, the dielectric constant
$\epsilon_{\rm F}$ is approximately $\epsilon_{\rm F} = 50
\epsilon_0$.\cite{CRC} The correlation function of the electrostatic
potential due to the charged impurities is:
\begin{equation}
\hat{\chi}_{\phi \phi}(k)=\frac{\hat{\chi}_{\rho \rho}(k)}{\epsilon^2 k^4}
\end{equation}
where now the dielectric constant should be the value in the zeolitic
solid, i.e. $\epsilon=4.2\epsilon_0$.\cite{CRC}

We first consider the diffusion of an ion of charge $q$ in such a
zeolite.  The energy-energy correlation function due to the charged
disorder is
\begin{equation}
\hat{\chi}_{vv}^{\rm ion}(k)=q^2 \hat{\chi}_{\phi \phi}(k)
\end{equation}
We consider diffusion of a singly charged species, $q=|e^-|$, at room
temperature. We take zeolite NaY with $\rho_{\rm T}=1.27\times
10^{28}\,$m$^{-3}$ and $\rho_{\rm Al}=0.3\,\rho_{\rm T}$ as our
example, where $\rho_{\rm T}$ is the density of T (=Si or Al) atoms in
the zeolite. We are interested in the magnitude of the time-dependent
diffusivity as well as the characteristic time, $t_{\rm c}$. It
appears at first sight that our theory requires no cutoff. Directly
solving eq \ref{D_infty} and \ref{matching}, we find
$D_{\infty}/D_0=e^{-634}$, and $D_0 \kappa^2 t_{\rm c}=8.8$. The term
$\beta^2 \chi(0)/d$ in the exponential of eq \ref{D_infty} looks like
an activation energy introduced by disorder, but the numerical result,
$\beta E=634$, is significantly larger than that measured in
experiments.\cite{Barrer}

It is, however, clear that Debye-H\"{u}ckel theory is not quite right
for short distances or large wave vectors. In particular, the
characteristic length of the charged disorder, $2\pi/\kappa$, is
typically on the order of several \AA{ngstroms}. Motion on the
\AA{ngstroms} length scale is measured in microscopic experiments such
as NMR and is well described by Newton's equation, e.g. by molecular
dynamics simulations. Statistical field theory is applicable only
for behavior at much longer time or length scales, and so $D_0$ is an
input parameter, determined from a simulation or a short-time
experiment. Since this $D_0$ already captures short time and distance
properties, a cutoff is required to exclude these effects. This cutoff
will be a function of $\kappa$. Given the measured diffusivity and the
experimental time scale, $t_{\rm exp}$, however, the cutoff can be
calculated by
\begin{eqnarray}
\Lambda_{\rm exp} &=& 2\pi/r_{\rm exp} \nonumber \\ 
r^2_{\rm exp} &=& 2D_0 t_{\rm exp}
\label{rexp}
\end{eqnarray}
Physically this means that we should use field theory only to consider
fluctuations with wavelengths larger than $r_{\rm exp}$.

Many factors, such as charge neutrality, Loewenstein's rule, strong
cation hydration, ion-sieving effects, and the existence of several
different energetically favorable cation sites, all complicate the
cationic diffusion process and make ion diffusion data hard to
organize.\cite{Karger} In Table \ref{ion1}, we illustrate the effect
of disorder for a few typical diffusivities. Since the available
techniques for measuring cation diffusion in zeolites, such as tracer
exchange or counter diffusion exchange, are at the time scale of
minutes, we use $t_{\rm exp}=100\,$s. From Table \ref{ion1}, we see
that a smaller $D_0$ leads to a larger cutoff and a larger correction
to the diffusivity. For example, if $D_0 = 1\times
10^{-10}\,$m$^2$s$^{-1}$ we see $D_{\infty}/D_0=1$ and no correction,
while with $D_0= 1\times 10^{-17}\,$m$^2$s$^{-1}$ we see a significant
correction of $D_{\infty}/D_0=0.002$. For all these cases, the cutoff
$\Lambda_{\rm exp}$ obtained is much smaller than $\kappa$, and so the
field-theoretic approach is appropriate. By assuming $\Lambda \ll
\kappa$, we can further simplify the energy-energy correlation
function as
\begin{equation}
\hat{\chi}_{vv}(k) = \frac{\eta}{k^2}; ~\mbox{with} ~\eta=\frac{q^2
\epsilon_{\rm F}} {\beta_{\rm F} \epsilon^2}
\end{equation}
From this form of $\hat{\chi}$, we find
\begin{equation}
D_{\infty}=D_0\exp[-\beta^2 K_d \eta \Lambda_{\rm exp}/d] \label{Deta}
\end{equation}
Solving for $t_{\rm c}$ from eq~\ref{matching}, we find
\begin{equation}
\Lambda_{\rm exp}^2 D_0 t_{\rm c} = 37.3
\end{equation}
and from eq \ref{mrg}, we find
\begin{equation}
t_0=t_{\rm c}/4
\end{equation}
From eq \ref{rexp} we see immediately that $t_{\rm c} \propto t_{\rm
exp}$, which explains why we would have the same value of $t_{\rm c}$
for the different diffusivities in Table \ref{ion1}. When the
diffusivity is dramatically reduced, the characteristic time is better
defined as the time at which the measured diffusivity is very close to
$D_{\infty}$, i.e. $D(t_{90\%})=1.1D_{\infty}$. We have calculated
$t_{90\%}$ numerically from eqs \ref{DRG} and \ref{tm} and list the values
in Table \ref{ion1}. As expected, $t_{90\%}$ increases with decreasing
$D_{\infty}/D_0$. Interestingly, we find the mean-square distance that
ions diffuse within this time period is constant:
\begin{equation}
\Delta x_{90\%}= {[R^2(t_{90\%})/3]}^{1/2}=\frac{\beta^2 \eta }{3\pi \ln(1.1)}
\left[\frac{t_0}{t_{\rm exp}}\right]^{1/2}
\end{equation}

We now turn to consider the case of diffusion of polarizable species
in zeolites. The energy of a polarizable species in an electric field
is given by $v({\bf x})=-\frac{1}{2} \alpha |\nabla \phi({\bf
x})|^2$, where $\alpha$ is the molecular polarizability, and $\phi$ is
the electric potential. Using the Gaussian approximation for $\phi$, we
find the energy-energy correlation function of polarizable species to
be
\begin{equation}
\hat{\chi}_{vv}^{\rm polarizable}({\bf k}) = \frac{1}{2}\alpha^2
\int_{\bf p} [({\bf p}-{\bf k}) \cdot {\bf p}]^2 \hat{\chi}_{\phi \phi}
({\bf p}-{\bf k}) \hat{\chi}_{\phi \phi} ({\bf p})
\label{pol}
\end{equation}
A cut-off is necessary to ensure the convergence of this integral. As
before, this cutoff must be smaller than $\kappa$ if the field theory is
to be applicable. Furthermore, the length scale should also be on the
order of or larger than a molecular size and roughly equal to the
displacement of molecules on the microscopic experimental time scale. We
simply set $\Lambda=2\pi/(5\mbox{\AA})$.

We list results of the calculation for the four species H$_2$, CH$_4$,
C$_4$H$_{10}$, and C$_6$H$_6$ in Table \ref{p1}. For H$_2$, the
transport and self-diffusion coefficients were measured simultaneously
in Ref~\citen{Jobic} by QENS. No discrepancy was found, and the
diffusivities given by the two methods are in good agreement with each
other, $D_0=3\times 10^{-9}\,$m$^2$s$^{-1}$ at $T=100\,$K in
NaX. Methane serves as another example where no discrepancies have
been reported. The diffusivity of methane in NaX measured by QENS is
$3.2 \times 10^{-9}\,$m$^2$s$^{-1}$.\cite{Jobic2} Our calculations
support the lack of discrepancy in these cases. Although the
characteristic time scales lie just within the range of QENS
experiments, the calculated $D_{\infty}/D_0$ for H$_2$ and
CH$_4$ is 0.72 and 0.39 respectively. Such a ratio would be difficult
to observe, given the small characteristic times.

Numerous experiments have reported that there are two orders of
magnitude of discrepancy between the microscopic and macroscopic
diffusion of benzene in NaX or NaY.\cite{Karger:p515} Computer
simulations by Auerbach {\it et al.} of benzene diffusion in Na-X or
NaY suggest that the longer time scale PFG-NMR technique measures
intercage diffusion, while the short time scale NMR spin-lattice
relaxation technique measures intracage diffusion.\cite{Auerbach,
Auerbach96, Auerbach2} These simulations also suggest that the
cage-to-cage length, 11\,\AA, is a reasonable approximation for the
effective hopping length. Our cutoff should be smaller than this value
to capture the effects of disorder on this length scale. We use the
microscopic $D_0=1.9 \times 10^{-10}\,$m$^2$s$^{-1}$ measured by
the QENS method.\cite{Jobic3} With our chosen cutoff, our theory
predicts $t_{90\%}=4.4\times 10^{-8}\,$s and
$D_{\infty}/D_0=0.08$. The large characteristic time is primarily due
to the smaller diffusivity for benzene compared to hydrogen or
methane. For this species, the greater discrepancy in diffusivity is
primarily due to the greater polarizability of benzene compared to
hydrogen or methane. For this species, we may expect to see the
time-dependence of the diffusivity in QENS experiments, or we may
expect to see discrepancies between microscopic and macroscopic
measurements.\cite{Karger} Consistent with the idea that increased
disorder slows down the diffusion species, benzene diffusivities observed
in simulations also decrease with increased aluminum
loading.\cite{Auerbach2} For butane, the calculated $D_{\infty}/D_0$
is small. Since the characteristic time $t_{90\%}$ lies within
the range of QENS, we may be able to observe a discrepancy in this
case as well.

Instead of imposing a length-scale cutoff for diffusion of the
polarizable species, we may impose a time-scale cutoff, as we did in
Table \ref{ion1}.  We choose $t_{\rm exp} = 10^{-9}$ s, since this is
at the lower end of the time scale accessible to QENS and near the upper
end of the time scale accessible to molecular dynamics.  We use eq
\ref{rexp} to derive the value of $\Lambda_{\rm exp}$ to use in our
calculations. The results are listed in Table \ref{p2}.  With this
type of cutoff, we find a discrepancy between microscopic and
macroscopic measurements only for benzene.  Comparison of the results
in Tables \ref{p1} and \ref{p2} shows that differences between
microscopic and macroscopic measurements depend sensitively on the
details of the time or length scale resolved by the microscopic experimental
technique. Experimental diffusivities may be less sensitive to the
cutoff procedure than these tables would suggest, since the Gaussian
approximation that we have used has lead to an over-sensitivity to
short-distance features in the case of polarizable species.

On the other hand, diffusion data for many species in zeolite A are in
satisfactory agreement.\cite{Eic} This seems at odds with our theory,
since in zeolite A, the Si:Al ratio is as high as unity. However, a
closer examination of the structure of zeolite A reveals a perfect
crystal structure. The aluminum ions obey Loewenstein's rule and are
arranged in a highly symmetrical way, such that each unit cell
consists of 96 AlO$_4$ and 96 SiO$_4$ tetrahedra with each apical
oxygen atom shared between one aluminum atom and one silicon
atom. Zeolite A is not disordered, and so the movement of
molecules within each unit cell rapidly reaches diffusive
behavior. We, therefore, do not expect to observe any long time
corrections to diffusivities in zeolite A. Although the Si:Al ratio is
close to unity in zeolite X as well, the unit cell is significantly
larger and the electrostatic potential is substantially disordered
compared to zeolite A, as judged by the presence of the strong
electric field gradients that make X-type zeolites suitable for
selective separation of air.\cite{Newsam} It is this disorder that can
lead to a time-dependent diffusivity.

We can summarize our results by noting that the characteristic time
for the slowing down of the diffusing species due to disorder is given
by $t_{\rm c} = \xi^2 /(2 D_0)$, where $\xi$ is a characteristic
length associated with the disorder. Smaller values of $D_0$,
therefore, lead to larger characteristic times and a greater chance
that a discrepancy between microscopic and macroscopic measurements
will occur.  We have considered only two contributions to the force on
a diffusing species, the Coulombic force on ions and the
induced-dipole force on polarizable species. Other interactions
between the diffusing species and the zeolite framework and
non-framework atoms, such as ion-dipole, ion-quadrapole,
Lennard-Jones, and so on, also contribute to the random energy fields.
These additional terms would also slow down the diffusing species,
although the characteristic times of most of these interactions are
too short to cause an observable discrepancy between microscopic and
macroscopic measurements.  Our calculations for diffusion of ions
should be fairly robust, since in this case there is especially good
reason to believe that the disorder is well-approximated by a
Gaussian.\cite{Deem} For polarizable species, our calculated
corrections to the short-time diffusivity are sensitive to the
numerical value of the cutoff. Equivalently, the theory is sensitive
to the assumptions about the behavior of the dielectric constant at
short distances. We have also assumed that the time scales associated
with configurational arrangement of the diffusing, polarizable species
are much shorter than the characteristic diffusion time induced by the
disorder. Our predictions for the ionic case, therefore, are more
quantitatively reliable than those for the polarizable
case. Regardless of the details, that disorder induces a
time-dependent diffusivity, as in eq ref{eq2}, for an ionic or polarizable species is a
general result.

Disorder on longer mesoscopic length scales can also occur in zeolites in the
form of faulting or growth-induced domain-mismatch. The length scales
associated with such domains can be on the order of $\xi = 10$ nm in
the ZSM-5\cite{Davis} and Linde Type L\cite{Tsapatsis} zeolites.  It
would be interesting to see if a reduction in the diffusivity
occurring at $t_{\rm c} = \xi^2/(2 D)$ is observable experimentally
for these materials.

\section{Conclusion}
\label{sec:last}
We have derived a time-dependent theory of diffusivity in disordered
microporous media. As judged by computer simulations, the theory is
reasonably accurate for Gaussian-type disorder over the entire time
range.  Applying the theory to zeolites, we offer disorder as one
possible mechanism for the existence of a discrepancy between
microscopic and macroscopic measurements of diffusion in zeolites.
For diffusion of ionic species, we predict that a discrepancy almost
certainly exists.  For polarizable species such as H$_2$ and CH$_4$ we
predict no observable discrepancy, and none has been observed
experimentally.  Our study of benzene and n-butane suggests that
disorder may cause a discrepancy for these molecules.  The
characteristic time and magnitude of the discrepancy are consistent
with microscopic and macroscopic experimental observations.  We
explain the lack of a discrepancy for zeolite A as resulting from a
lack of disorder in this highly-symmetric, perfectly-crystalline
material.

Our theory can only predict the ratio $D(t)/D_0$.  Determination of
the short-time diffusivity, $D_0$, is best done by a molecular
dynamics simulation or a microscopic experiment. In this respect,
molecular dynamics simulation is a particularly useful, complementary
tool to the field theory presented here. Diffusivities derived by
molecular dynamics are often comparable with the NMR, PFG NMR, and
QENS measurements.\cite{Snurr} Moreover, simulation provides molecular
details beyond the scope of an experiment or a field theory. Such
simulations should, for example, allow one to construct a more exact
form of $\hat{\chi}_{vv}(k)$ to be used in our theory.

\section*{Acknowledgments}
We thank Jeong-Man Park for useful discussions.  This research was
supported by Chevron Research and Technology Company and by the
National Science Foundation through grant number CTS--9702403.

\bibliography{diffuse}

\providecommand{\refin}[1]{\\ \textbf{Referenced in:} #1}
\begin{thebibliography}{10}

\bibitem{Karger}
K\"{a}rger,~J.;\ \ Ruthven,~D.~M. \textit{Diffusion in Zeolites and other
  Microporous Solids;} John Wiley \& Sons, Inc: New York, 1991 especially
  Chapter 15.

\bibitem{Eic}
Eic,~M.;\ \ Ruthven,~D.~M. \textit{Zeolites} \textbf{1988,} \textsl{8,}
  472-479.

\bibitem{Sun}
Sun,~M.~S.;\ \ Talu,~O.;\ \ Shah,~D.~B. \textit{AIChE J.} \textbf{1996,}
  \textsl{42,} 3001-3007.

\bibitem{Karger2}
K\"{a}rger,~J.;\ \ Pfeifer,~H.;\ \ Rauscher,~M.;\ \ Walter,~A. \textit{J.C.S.
  Faraday I} \textbf{1980,} \textsl{76,} 717-737.

\bibitem{Jobic}
Jobic,~H.;\ \ K\"{a}rger,~J.;\ \ B\'{e}e,~M. \textit{Phys. Rev. Lett.}
  \textbf{1999,} \textsl{82,} 4260-4263.

\bibitem{Karger:p515}
Ref.\ \citen{Karger}, p.\ 515.

\bibitem{Runnebaum}
Runnebaum,~R.~C.;\ \ Maginn,~E.~J. \textit{J. Phys. Chem. B} \textbf{1997,}
  \textsl{101,} 6394-6408.

\bibitem{Auerbach}
Auerbach,~S.~M.;\ \ Henson,~N.~J.;\ \ Cheetham,~A.~K.;\ \ Metiu,~H.~I.
  \textit{J. Phys. Chem.} \textbf{1995,} \textsl{99,} 10600-10608.

\bibitem{Newsam}
Newsam,~J.~M.  Zeolites.   In  \textit{Solid State Chemistry: Compounds};
  Cheetham,~A.~K.;\ \ Day,~P.,\ \ Eds.;  Oxford University Press: Oxford, 1992.

\bibitem{Deem}
Deem,~M.~W.;\ \ Chandler,~D. \textit{J. Stat. Phys.} \textbf{1994,}
  \textsl{76,} 911--927.

\bibitem{Auerbach2}
Auerbach,~S.~M.;\ \ Metiu,~H.~I. \textit{J. Chem. Phys.} \textbf{1997,}
  \textsl{106,} 2893--2905.

\bibitem{Deem3}
Park,~J.-M.;\ \ Deem,~M.~W. \textit{Phys. Rev. E} \textbf{1998,} \textsl{58,}
  1487--1493.

\bibitem{Deem4}
Deem,~M.~W. \textit{Phys. Rev. Lett.} \textbf{1999,} \textsl{83,} 1694.

\bibitem{Deem2}
Deem,~M.~W.;\ \ Park,~J.-M. \textit{Phys. Rev. E} \textbf{1998,} \textsl{57,}
  2681-2685.

\bibitem{Park}
Park,~J.-M.;\ \ Deem,~M.~W. \textit{Phys. Rev. E} \textbf{1998,} \textsl{57,}
  3618-3621.

\bibitem{Justin}
Zinn-Justin,~J. \textit{Quantum Field Theory and Critical Phenomena;} Clarendon
  Press: Oxford, 3rd ed.; 1996 {C}hapter 11.

\bibitem{Dean}
Dean,~D.~S.;\ \ Drumond,~I.~T.;\ \ Horgan,~R.~R. \textit{J. Phys. A: Math.
  Gen.} \textbf{1994,} \textsl{27,} 5135--5144 Figure 8.

\bibitem{Victor}
Pham,~V.;\ \ Deem,~M.~W. \textit{J. Phys. A: Math. Gen.} \textbf{1998,}
  \textsl{31,} 7235-7247.

\bibitem{Ma}
Ma,~S.~K. \textit{Modern Theory of Critical Phenomena;} volume~46 of
  \textit{Fronteiers in Physics} The Benjamin/Cummings Publishing Company, Inc:
  London, 1976 Chapter X.

\bibitem{Netz}
Netz,~R.~R.;\ \ Orland,~H. \textit{Europhys. Lett.} \textbf{1999,} \textsl{45,}
  726-732.

\bibitem{CRC}
Lide,~D.~R.,\ \ Ed.;  \textit{CRC handbook of Chemistry and Physics;} CRC
  Press: New York, 80th ed.; 1999.

\bibitem{Barrer}
Barrer,~R.~M.  Zeolite Exchangers-Some Equilibrium and Kinetic Aspects.   In
  \textit{Proceedings of the fifth international conference on zeolites};
  Rees,~L. V.~C.,\ \ Ed.;  Heyden: London, 1980 Table 3.

\bibitem{Jobic2}
Jobic,~H.;\ \ B\'{e}e,~M.;\ \ Kearley,~G.~J. \textit{J. Phys. Chem.}
  \textbf{1994,} \textsl{98,} 4660-4665.

\bibitem{Auerbach96}
Auerbach,~S.~M.;\ \ Bull,~L.~M.;\ \ Henson,~N.~J.;\ \ Metiu,~H.~I.;\ \
  Cheetham,~A.~K. \textit{J. Phys. Chem.} \textbf{1996,} \textsl{100,}
  5923-5930.

\bibitem{Jobic3}
Jobic,~H.;\ \ B\'{e}e,~M.;\ \ K\"{a}rger,~J.;\ \ Pfeifer,~H.;\ \ Caro,~J.
  \textit{J. Chem. Soc., Chem. Comm.} \textbf{1990,}  341--342.

\bibitem{Davis}
de~Moor,~P. E.~A.;\ \ Beelen,~T. P.~M.;\ \ Komanschek,~B.~U.;\ \ Beck,~L.;\ \
  Wagner,~P.;\ \ Davis,~M.;\ \ van Santen,~R. \textit{Chem. Eur. J.}
  \textbf{1999,} \textsl{5,} 2083-2088 Figure 3.

\bibitem{Tsapatsis}
Nikolakis,~V.;\ \ Vlachos,~D.~G.;\ \ Tsapatsis,~M. \textit{J. Chem. Phys.}
  \textbf{1999,} \textsl{111,} 2143--2150.

\bibitem{Snurr}
Clark,~L.~A.;\ \ Ye,~G.~T.;\ \ Gupta,~A.;\ \ Hall,~L.~L.;\ \ Snurr,~R.~Q.
  \textit{J. Chem. Phys.} \textbf{1999,} \textsl{111,} 1209-1222.

\end{thebibliography}

\clearpage
\newpage

\begin{table}
\centering
\begin{minipage}[htbp]{6.00in}
\renewcommand{\footnoterule}{}
\caption{
Measurements of self diffusivities and activation energies in
zeolite Y with different Al content.\cite{Barrer} }
\begin{tabular}{ccclcc} \hline
Zeolite& Si:Al &
$\rho_{\rm Al}/\rho_{\rm T}$\footnote{$\rho_{\rm T}$ 
            is the density of framework T (=Si or Al) atoms.}
 &Solvent & E\,(kJ/mol)&
D$\,($m$^2$s$^{-1})$\\ \hline

Sr-Y & 139:53 & 0.276 &Methanol & 91.6 & $5.2\times 10^{-22}$ \\ \hline 

Sr-Y & 125:67 & 0.349 &Methanol & 118 & $1.4\times 10^{-25}$ \\ \hline

Ba-Y & 139:53 & 0.276 &Methanol & 67.8 & $3.3\times 10^{-20}$ \\
\hline

Ba-Y & 125:67 & 0.349 &Methanol & 84.1 & $1.5\times 10^{-21}$ \\
\hline

Ba-Y & 139:53 & 0.276 &Ethanol & 91.6 & $3.3\times 10^{-24}$ \\ \hline

Ba-Y & 125:67 & 0.349 &Ethanol & 120 & $3.1\times 10^{-25}$ \\ \hline
\end{tabular}
\label{ions}
\end{minipage}
\end{table}

\begin{table}
\begin{center}
\caption{
Ionic diffusivities in NaY with $\rho_{\rm Al}=0.3\,\rho_{\rm
T}$ and $T=298\,$K.
% \epsilon_{\rm F}=50
}
\begin{tabular}{cccccc} \hline
$D_0\,($m$^2$s$^{-1})$ & $t_{\rm exp}\,$(s) & $\Lambda_{\rm exp}\,($m$^{-1})$ & $D_{\infty}/D_0$ & $t_{90\%}\,$(s) & $\Delta x_{90\%}\,$(m) \\ \hline

$1 \times 10^{-10}$  & 100  & $4.4 \times 10^{4}$  &1 &- & -\\ \hline

$1 \times 10^{-14}$  & 100  & $4.4 \times 10^{6}$  &0.82 &230 &
$2.03\times 10^{-6}$ \\ \hline

$1 \times 10^{-15}$  & 100  & $1.4 \times 10^{7}$  &0.53 &$3.5\times
10^{3}$ & $2.03\times 10^{-6}$\\ \hline

$1 \times 10^{-16}$  & 100  & $4.4 \times 10^{7}$  &0.14 &$1.4\times
10^{5}$ & $2.03\times 10^{-6}$\\ \hline

$1 \times 10^{-17}$  & 100  & $1.4 \times 10^{8}$  &0.002 &$1.0\times
10^{8}$&$2.03\times 10^{-6}$ \\ \hline

$1 \times 10^{-18}$  & 100  & $4.4 \times 10^{8}$  &$e^{-20}$
&$8.6\times 10^{14}$ & $2.03\times 10^{-6}$\\ \hline

\end{tabular}
\label{ion1}
\end{center}
\end{table}

\small

\begin{table}
\begin{minipage}[htbp]{6.00in}
\renewcommand{\footnoterule}{}
\caption{}
Characteristic time, length, and $D_{\infty}/D_0$ for polarizable
species with $\Lambda=2\pi/(5\mbox{\AA})$.\protect\footnote{If a
short-distance dielectric constant of $\epsilon=2.0\epsilon_0$ were
used, a cutoff of $\Lambda=2\pi/(9.5\mbox{\AA})$ would lead to similar
values of $D_{\infty}/D_0$ and characteristic times increased by
roughly a factor of three.} 

\begin{tabular}{cccccccccc} \hline
Species &Zeolite & $\rho_{\rm Al}/\rho_{\rm T}$
&T(K)&$D_0\,$(m$^2$s$^{-1}$) &Ref& $\alpha/(4\pi\epsilon_0)\,(
$m$^{3})$\cite{CRC} & $D_{\infty}/D_0$ & $t_{90\%}\,$(s)

&$\Delta x_{90\%}\mbox{(\AA)}$\\ \hline

H$_2$ &NaX &0.5 &100 & $3\times 10^{-9}$&\citen{Jobic}&$0.8\times
10^{-30}$& 0.72  &$7.1\times 10^{-11}$ &5.8 \\ \hline 

CH$_4$ &NaY &0.3 &150 &$3.2 \times 10^{-9}$ &\citen{Jobic2}&
$2.6\times 10^{-30}$ & 0.39  &$3.0\times 10^{-10}$ &9.1\\ \hline

C$_4$H$_{10}$ & NaX & 0.5&298& $2.5\times 10^{-9}$&\citen{Karger:p515}
&$8.2\times 10^{-30}$& 0.02 &$2.0\times 10^{-8}$ & 15\\ \hline

C$_6$H$_6$ &NaX &0.5 &458 &$1.9\times 10^{-10}$&\citen{Jobic3} & $10
\times 10^{-30}$ & 0.08 &$4.5\times 10^{-8}$ & 13\\ \hline

\end{tabular}
\label{p1}
\end{minipage}
\end{table}

\begin{table}
\begin{minipage}[htbp]{6.00in}
\renewcommand{\footnoterule}{}
\caption{}
Characteristic time, length, and $D_{\infty}/D_0$ for polarizable
species with $t_{\rm exp}=10^{-9}$\,s.\protect\footnote{If a
short-distance dielectric constant of $\epsilon=2.0\epsilon_0$ were
used, a $t_{\rm exp}=3.5\times 10^{-9}$s would lead to similar value of
$D_{\infty}/D_0$ and a characteristic time increased to $1.9 \times
10^{-8}$s for benzene}
\begin{tabular}{cccccccccc} \hline
Species &Zeolite & $\rho_{\rm Al}/\rho_{\rm T}$
&T(K)&$D_0\,$(m$^2$s$^{-1}$) &Ref& $\alpha/(4\pi\epsilon_0)\,(
$m$^{3})$\cite{CRC} & $D_{\infty}/D_0$ & $t_{90\%}\,$(s)
& $\Delta x_{90\%}\mbox{(\AA)}$\\ \hline

H$_2$ &NaX &0.5 &100 & $3\times 10^{-9}$ &\citen{Jobic}&$0.8\times
10^{-30}$& 1  &- &-\\ \hline 

CH$_4$ &NaY &0.3 &150 &$3.2 \times 10^{-9}$ &\citen{Jobic2}&
$2.6\times 10^{-30}$ & 1  &- &-\\ \hline

C$_4$H$_{10}$ & NaX & 0.5&298& $2.5\times 10^{-9}$&\citen{Karger:p515}
&$8.2\times 10^{-30}$& 0.999 &- &-\\ \hline

C$_6$H$_6$ &NaX &0.5 &458 &$1.9\times 10^{-10}$&\citen{Jobic3} & $10
\times 10^{-30}$ & 0.37 &$7.5\times 10^{-9}$ &11\\ \hline

\end{tabular}
\label{p2}
\end{minipage}
\end{table}

\normalsize
\clearpage
\newpage



\end{document}